\documentclass[9pt,twocolumn,twoside]{optica}
\setboolean{shortarticle}{false}
\setboolean{minireview}{false}
 \usepackage{multirow}
 \usepackage{url}

\title{Imaging through glass diffusers using densely connected convolutional networks  }

\author[1,*]{Shuai Li}
\author[2]{Mo Deng}
\author[3]{Justin Lee}
\author[1,\P]{Ayan Sinha}
\author[1,4]{George Barbastathis}

\affil[1]{Department of Mechanical Engineering, Massachusetts Institute of Technology, 77 Massachusetts Avenue, Cambridge, MA 02139}
\affil[2]{Department of Electrical Engineering and Computer Science, Massachusetts Institute of Technology, 77 Massachusetts Avenue, Cambridge, MA 02139}
\affil[3]{Institute for Medical Engineering Science, Massachusetts Institute of Technology, 77 Massachusetts Avenue, Cambridge, MA 02139}
\affil[4]{Singapore-MIT Alliance for Research and Technology (SMART) Centre, One Create Way, Singapore 117543, Singapore}

\affil[*]{Corresponding author: shuaili@mit.edu}

\affil[$\P$]{Current address:  Magic Leap Inc,1376 Bordeaux Drive, Sunnyvale, CA 94089 }

\ociscodes{(100.3190) Inverse problems; (100.4996) Pattern recognition, neural networks; (110.1758) Computational imaging.}

\begin{abstract}
Computational imaging through scatter generally is accomplished by first characterizing the scattering medium so that its forward operator is obtained; and then imposing additional priors in the form of regularizers on the reconstruction functional so as to improve the condition of the originally ill-posed inverse problem. In the functional, the forward operator and regularizer must be entered explicitly or parametrically (e.g. scattering matrices and dictionaries, respectively.) However, the process of determining these representations is often incomplete, prone to errors, or infeasible. Recently, deep learning architectures have been proposed to instead learn both the forward operator and regularizer through examples. Here, we propose for the first time, to our knowledge, a convolutional neural network architecture called ``IDiffNet'' for the problem of imaging through diffuse media and demonstrate that IDiffNet has superior generalization capability through extensive tests with well-calibrated diffusers. We found that the Negative Pearson Correlation Coefficient loss function for training is more appropriate for spatially sparse objects and strong scattering conditions. Our results show that the convolutional architecture is robust to the choice of prior, as demonstrated by the use of multiple training and testing object databases, and capable of achieving higher space-bandwidth product reconstructions than previously reported. 
\end{abstract}

\setboolean{displaycopyright}{true}

\begin{document}

\maketitle

\section{Introduction} \label{sec:intro}
Imaging through random media \cite{tatarski2016wave, ishimaru1978wave} remains one of the most useful as well as challenging topics in computational optics. This is because scattering impedes information extraction from the wavefront in two distinct, albeit related ways. First, light scattered at angles outside the system's Numerical Aperture is lost; second, the relative phases among spatial frequencies that pass are scrambled---convolved with the diffuser's own response. In most cases, the random medium is not known or it is unaffordable to characterize it completely. Even if the random medium and, hence, the convolution kernel are known entirely, deconvolution is highly ill-posed and prone to noise-induced artifacts. 

Therefore, the strategy to recover the information, to the degree possible, must be two-pronged: first, to characterize the medium as well as possible so that at least errors in the deconvolution due to incomplete knowledge of the medium's response may be mitigated; and, second, to exploit additional {\it a priori} knowledge about the class of objects being imaged so that the inverse problem's solution space is reduced and  spurious solutions are excluded. These two strategies are summarized by the well-known Tikhonov-Wiener optimization functional for solving inverse problems as
\begin{equation}
\hat{f} = \text{argmin}_f \left\{\rule[-1ex]{0cm}{3ex} \left|\!\left| g-Hf \right|\!\right|^2 + \alpha \Phi(f) \right\},
\end{equation}
where $H$ is the forward operator in the optical system $g=Hf$, $f$ is the unknown object, $g$ is the raw intensity image (or images if some form of scanning is involved), $\Phi(.)$ expresses prior knowledge by penalizing unacceptable objects so the optimization is prohibited from landing onto them, $\alpha$ is the regularization parameter controlling the relative contribution of the two terms in the functional, and $\hat{f}$ is the estimate of the object. 

The forward operator $H$ includes the effects of the scatterer, as well as of the optical system utilized in any particular situation. A number of ingenious strategies have been devised to design forward operators that improve the imaging problem's condition, most famously by using nonlinear optics \cite{denk1990two, moreaux2000membrane} or stimulated emission \cite{hell1994breaking}. Restricting oneself to linear optics, structured illumination \cite{gustafsson2000surpassing, wilson2011optical, lim2008wide} is an effective strategy which modulates object information onto better-behaved spatial frequencies.

Several approaches characterize the random medium efficiently. One method is to measure the transmission matrix of the medium by interferometry or wavefront sensing \cite{popoff2010measuring, popoff2010image, dremeau2015reference}. Alternatively, one may utilize the angular memory effect in speckle correlation \cite{bertolotti2012non, katz2014non, stasio2016calibration, porat2016widefield}. The angular memory principle states that rotating the incident beam over small angles does not change the resulting speckle pattern but only translates it over a small distance \cite{feng1988correlations, freund1988memory}. In this case, computing the autocorrelation of the output intensity and deconvolving it by the autocorrelation function of the speckles, which is a sharply peaked function \cite{akkermans2007mesoscopic}, results in the autocorrelation of the input field. Then, the object is recovered from its own autocorrelation using the Gerchberg-Saxton-Fienup (GSF) algorithm \cite{gerchberg1972practical, fienup1978reconstruction} with additional prior constraints.

Turning to the problem of determining $\Phi$, during the past two decades thanks to efforts by Grenander \cite{grenander1993general}, Cand\'es \cite{candes2006robust}, and Brady \cite{brady2009compressive}, the use of sparsity priors was popularized and proved to be effective in a number of contexts including random media. For example, Liu {\it et al} successfully recovered the 3D positions of multiple LEDs embedded in turbid scattering media by taking phase-space measurements and imposing the $\mathcal{L}1$ sparsity prior \cite{liu20153d}.

Instead of establishing $H$ and $\Phi$ independently and explicitly from measurements and prior knowledge, an alternative approach is to {\em learn} both operators simultaneously through examples of objects imaged through the random medium. To our knowledge, the first instance when this strategy was put forth was by Horisaki \cite{horisaki2016learning}. In that paper, a Support Vector Regression (SVR) learning architecture was used to learn the scatterer and the prior of faces being imaged through. The approach was effective in that the SVR learned correctly to reconstruct face objects; it also elucidated the generalization limitations of SVRs, which are shallow two-layer architectures, as for example when presented with non-face objects the SVR would still respond with one of its learned faces as a reconstruction. A deeper fully-connected architecture in the same learning scheme has been proposed recently \cite{lyu2017exploit}. The Horisaki paper was the first, to our knowledge, to use machine learning in the computational imaging context; it certainly influenced our own work on lensless imaging \cite{Sinha:17} and other related works  \cite{rivenson2017phase, rivenson2017deep, jin2017deep, satat2017object, horstmeyer2017convolutional}. 

In this paper, we propose for the first time, to our knowledge, two innovations in the use of machine learning for imaging through scatter: the first is the use of the convolutional network architecture \cite{lecun1995convolutional} and the second the use of Negative Pearson Correlation Coefficient (NPCC) as loss function. The convolutional architecture, which we refer to as IDiffNet (for ``Imaging through Diffusers Network''), affords us two main advantages: first, we are able to reconstruct space-bandwidth product (SBP) of $128\times 128$, higher than previously reported; second, IDiffNet is able to better generalize when reconstructing test objects from classes outside its own training set (e.g. reconstructing faces when trained on natural images.) We conducted training and testing with well-calibrated diffusers of known grit size and well-calibrated intensity objects produced by a spatial light modulator. We also examined a large set of databases, including classes of objects with naturally embedded sparsity (e.g. handwritten characters or digits). These experiments enabled us to precisely quantify when IDiffNet requires strong sparsity constraints to become effective, as function of diffuser severity (the smaller the grit size, the more ill-posed the inverse problem becomes.) 

The adoption of NPCC instead of the more commonly used Mean Absolute Error (MAE) as loss function for training IDiffNet was an additional enabling factor in obtaining high-SBP image reconstructions through strong scatter. We compared the performance of these two loss functions under different imaging conditions and with different training databases determining the object priors that the networks learn and showed that NPCC is preferable for cases of relatively sparse objects (e.g. characters) and strong scatter. Lastly, we probed the interior of our trained IDiffNets through the well-established test of Maximally-Activated Patterns (MAPs) \cite{zeiler2014visualizing} and compared with standard denoising networks to eliminate the possibility that IDiffNet might be acting trivially instead of having learnt anything about the diffuser and the objects' priors. 

The structure of the paper is as follows: In Section~\ref{sec:archi}, we describe the architecture of our computational imaging system, including the hardware and the IDiffNet machine learning architecture. In Section~3, the reconstruction results are analyzed, including the effects of scattering strength, object complexity ({\it i.e.}, the object priors that the neural networks must learn) and choice of loss function for training. The comparison with a denoising neural network is described in Section~4 and  concluding thoughts are  in Section~5.

\section{Computational imaging system architecture} \label{sec:archi}
\begin{figure}[h!]
\centering\includegraphics[width=1\linewidth]{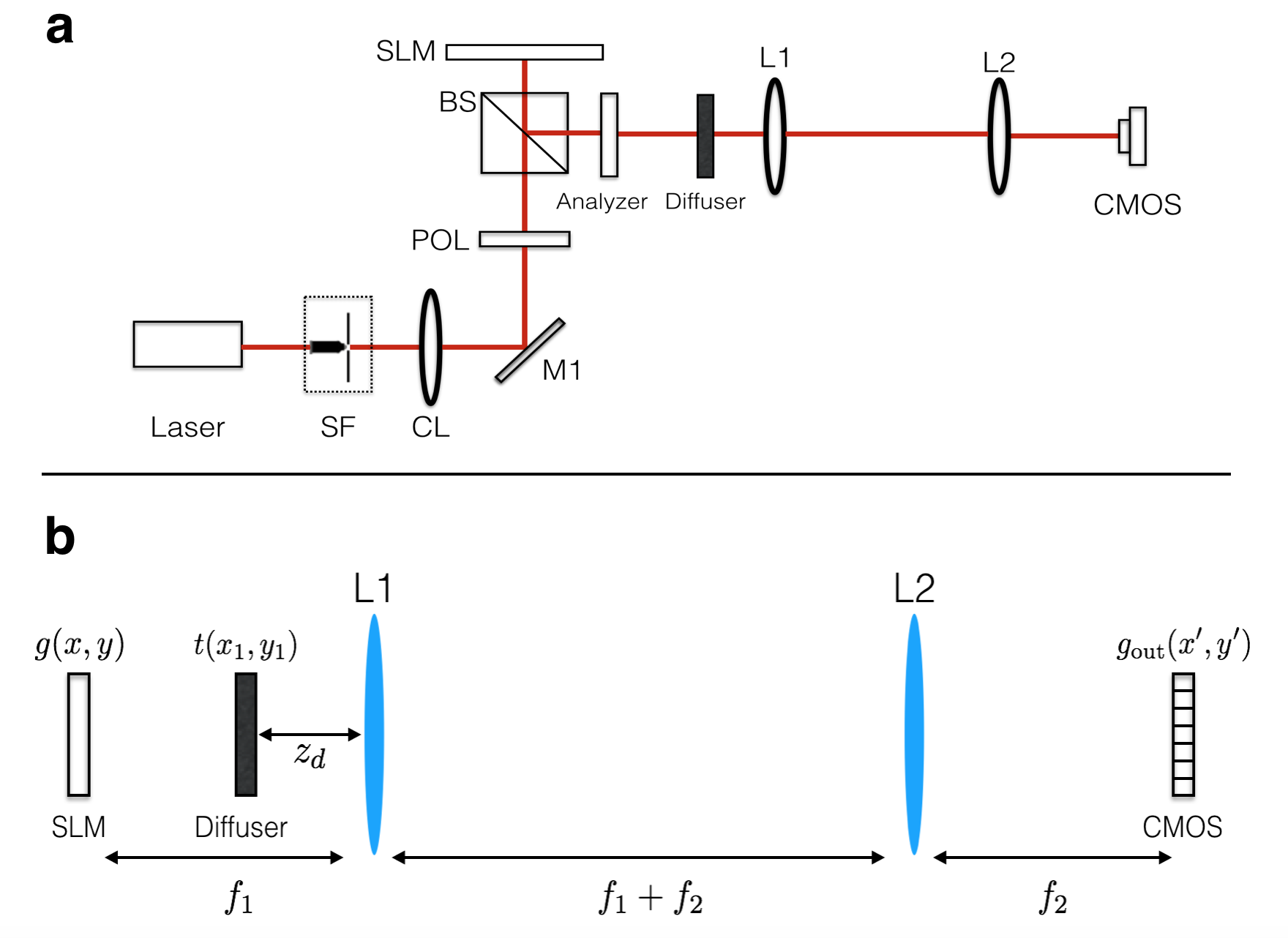}
\caption{Optical configuration. (a) Experimental arrangement. SF: spatial filter; CL: collimating lens; M: mirror; POL: linear polarizer; BS: beam splitter; SLM: spatial light modulator. (b) telescopic imaging system }
\label{fig:system}
\end{figure}

The optical configuration that we consider in this paper is shown in Fig. \ref{fig:system}. Light from a He-Ne laser source (Thorlabs, HNL210L, 632.8nm) is transmitted through a spatial filter, which consists of a microscope objective (Newport, M-60X, 0.85NA) and a pinhole aperture ($D=5\mu \text{m}$). After being collimated by the lens ($f=150$mm), the light is reflected by a mirror and then passes through a linear polarizer, followed by a beam splitter. A spatial light modulator (Holoeye, LC-R 720, reflective) is placed normally incident to the transmitted light and acts as a pixel-wise intensity object. The SLM pixel size is $20\times 20\mu\text{m}^2$ and number of pixels is $1280\times 768$, out of which the central $512\times 512$ portion only is used in the experiments. The SLM-modulated light is then reflected by the beam splitter and passes through a linear polarization analyzer before being scattered by a glass diffuser. A telescopic imaging system is built after the glass diffuser to image the SLM onto a CMOS camera (Basler, A504k), which has a pixel size of $12\times 12\mu\text{m}^2$. In order to match the pixel size of the CMOS with that of the SLM, we built the telescope using two lenses $L_{1}$ and $L_{2}$ of focal lengths: $f_{1}=250mm$ and $f_{2}=150mm$. As a result, the telescope magnifies the object by a factor of $0.6$, which is consistent with the ratio between the pixel sizes of the CMOS and SLM. The total number of pixels on the CMOS is $1280\times 1024$, but we only crop the central $512\times 512$ square for processing; thus, the number of pixels measured by the CMOS camera, as well as their size, match 1:1 the object pixels at the SLM. Images recorded by the CMOS camera are then processed on an Intel i7 CPU. The neural network computations are performed on a GTX1080 graphics card (NVIDIA).

The modulation performance of the SLM depends on the orientations of the polarizer and analyzer. Here, we implement the cross polarization arrangement to achieve a high intensity modulation contrast. Specifically, we set the incident beam to be linearly polarized along the horizontal direction and also set the linear polarization analyzer to be oriented along the vertical direction. We experimentally calibrate the correspondence between the 8-bit grayscale input images projected onto the SLM and intensity modulation values of SLM (see Supplement). We find that in this arrangement, the intensity modulation of the SLM follows a monotonic relationship with respect to assigned pixel value and a maximum intensity modulation ratio of $\sim 17$ can be achieved. At the same time, the SLM also introduces phase modulation which is correlated with the intensity modulation due to the optical anisotropy of the liquid crystal molecules. The phase depth is $\sim 0.6\pi$. Fortunately, the influence of this phase modulation is negligible in the formation of the speckle images that we captured in this system; detailed demonstration can be found in the Supplement. Therefore, we are justified in treating this SLM as a pure-intensity object.

As shown in Fig. \ref{fig:system}(b), the glass diffuser is inserted at a distance $z_{d}$ in front of the lens L1. Here, we approximate the glass diffuser as a thin mask whose amplitude transmittance is $t(x_{1}, y_{1})$. In this case, a forward model can be derived to relate the optical field at the detector plane $g_{\text{out}}(x',y')$ to the optical field at the object plane $g(x,y)$ (constant terms have been neglected) \cite{goodman2005introduction}:
\begin{equation}
\begin{aligned}
&g_{\text{out}}(x',y')=\left\{e^{\frac{-i\pi f_{1}^2}{\lambda(f_{1}-z_d)f_{2}^2}(x'^2+y'^2)}\cdot\int\int dxdy \left[g(x,y)e^{\frac{i\pi}{\lambda(f_{1}-z_d)}(x^2+y^2)}\cdot\right.\right.\\
&\left.\left.T\left(\frac{x+f_{1}x'/f_{2}}{\lambda(f_{1}-z_d)},\frac{y+f_{1}y'/f_{2}}{\lambda(f_{1}-z_d)}\right)\right]\right\}*\left[\frac{J_{1}(\frac{2\pi R}{\lambda f_{2}}\sqrt{x'^2+y'^2})}{\sqrt{x'^2+y'^2}}\right]\\
\end{aligned}
\label{eqs:1}
\end{equation}
where $\lambda$ is the light wavelength, $R$ the radius of the lens $L_{2}$ and $J_{1}(\cdot)$ denotes the first-order Bessel function of the first kind. $T$ is the Fourier spectrum of the diffuser: $T(u,v)=\int\int dx_{1}dy_{1} \left[t(x_{1}, y_{1})e^{-i2\pi(x_{1}u+y_{1}v)}\right]$. Here, $*$ denotes the convolution product and the last term in the convolution accounts for the influence of the finite aperture size of the lenses. 

We model the diffuser transmittance $t(x_{1},y_{1})$ as a pure-phase random mask, {\it i.e.} $t(x_{1},y_{1})=\exp\left[\frac{i2\pi\Delta n}{\lambda}D(x_{1},y_{1})\right]$, where $D(x_{1},y_{1})$ is the random height of the diffuser surface and $\Delta n$ is the difference between the refractive indices of the diffuser and the surrounding air ($\Delta n\approx0.52$ for glass diffusers). The random surface height $D(x_{1},y_{1})$ can be modeled as \cite{antipa2016single}:
\begin{equation}
D(x,y)=W(x,y)*K(\sigma).
\label{eqs:2}
\end{equation}
Here, $W(x,y)$ is a set of random height values chosen according to the normal distribution at each discrete sample location $(x,y)$, i.e. $W\sim N(\mu, \sigma_{0})$; and $K(\sigma)$ is a zero-mean Gaussian smoothing kernel having full-width half-maximum (FWHM) value of $\sigma$. 

\begin{figure}[h!]
\centering\includegraphics[width=1\linewidth]{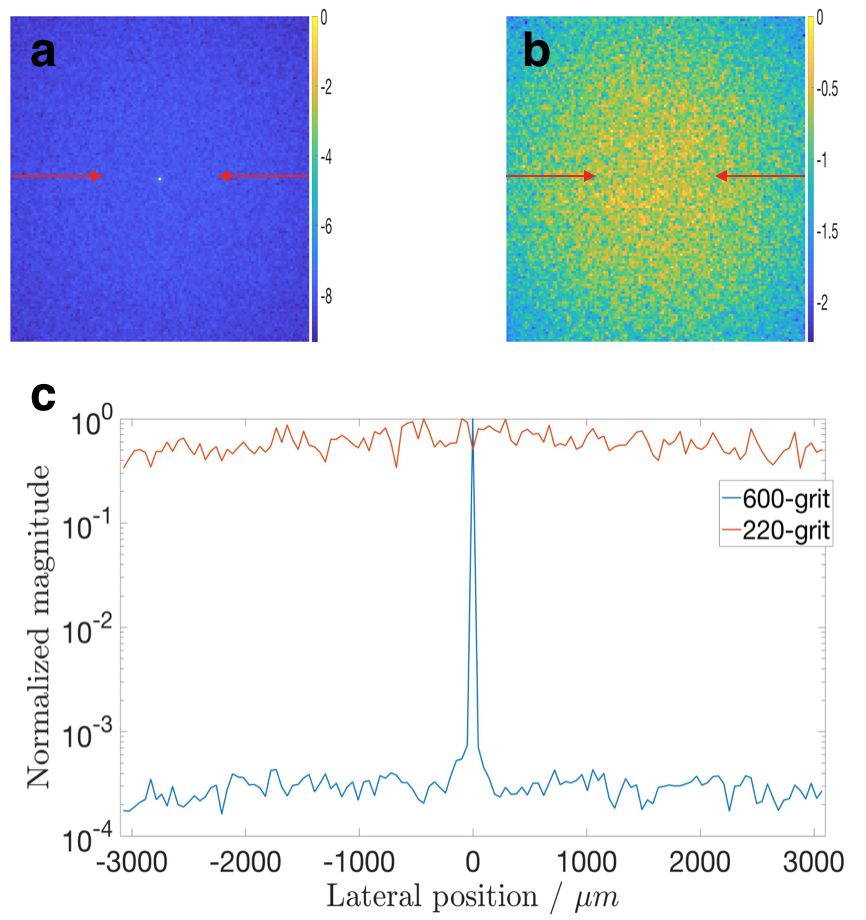}
\caption{Point spread functions (PSFs) of the system. (a) PSF for the 600-grit diffuser: $\mu=16\mu\text{m}$, $\sigma_{0}=5\mu\text{m}$, $\sigma=4\mu\text{m}$. (b) PSF for the 220-grit diffuser: $\mu=63\mu\text{m}$, $\sigma_{0}=14\mu\text{m}$, $\sigma=15.75\mu\text{m}$. (c) Comparison of the profiles of the two PSFs alone the lines indicated by the red arrows in (a) and (b). Other simulation parameters are set to be the same as the actual experiment: $z_{d}=15\text{mm}$, $R=12.7\text{mm}$ and $\lambda=632.8\text{nm}$.  All the plots are in logarithmic scale.}
\label{fig:psf}
\end{figure}

The values of $\mu$, $\sigma_{0}$ and $\sigma$ are determined by the grit size of the chosen glass diffuser \cite{grit}. In this paper, we use two glass diffusers of different grit size: 600-grit (Thorlabs, DG10-600-MD) and 220-grit (Edmund, 83-419). Using these values in (\ref{eqs:1}) and (\ref{eqs:2}), we simulate the point spread function (PSF) of our imaging system as shown in Fig. \ref{fig:psf}.  We can see that the PSF for the 600-grit diffuser has a sharp peak at the center, while the PSF for the 220-grit diffuser spreads more widely. This indicates that the 220-grit diffuser scatters the light much more strongly than the 600-grit diffuser.

We may also represent equation (\ref{eqs:1}) in terms of a forward operator $H_{g}$: $g_{\text{out}}(x',y')=H_{g}g(x,y)$. When the object is pure-intensity, {\it i.e.} $g(x,y)=\sqrt{I(x,y)}$, the relationship between the raw intensity captured at the detector plane $I_{\text{out}}(x',y')$ and the object intensity $I(x,y)$ can also be represented in terms of another forward operator $H$: $I_{\text{out}}(x',y')=HI(x,y)=[SH_{g}Sr]I(x,y)$. Here, $S$ denotes the modulus square operator and $Sr$ denotes the square root operator. Then, in order to reconstruct the intensity distribution of the object, we have to formulate an inverse operator $H^{\text{inv}}$ such that
\begin{equation}
\hat{I}(x,y)=H^{\text{inv}}I_{\text{out}}(x',y')
\end{equation}
\label{eqs:3}
where $\hat{I}(x,y)$ is an acceptable estimate of the intensity object.

Due to the randomness of $H$, it is difficult to obtain its explicit form and do the inversion accordingly; prior works referenced in Section~\ref{sec:intro} employed measurements of the scattering matrix to obtain $H$ approximately. Here, we instead use IDiffNet, a deep neural network (DNN) trained to the underlying inverse mapping given a set of training data. IDiffNet uses the densely connected convolutional architecture (DenseNets) \cite{huang2016densely}, where each layer connects to every other layer within the same block in a feed-forward fashion. Compared with conventional convolutional networks with same number of layers, DenseNets have more direct connections between the layers, which strengthens feature propagation and encourages feature reuse. 

\begin{figure*}[h!]
\centering\includegraphics[width=0.6\linewidth]{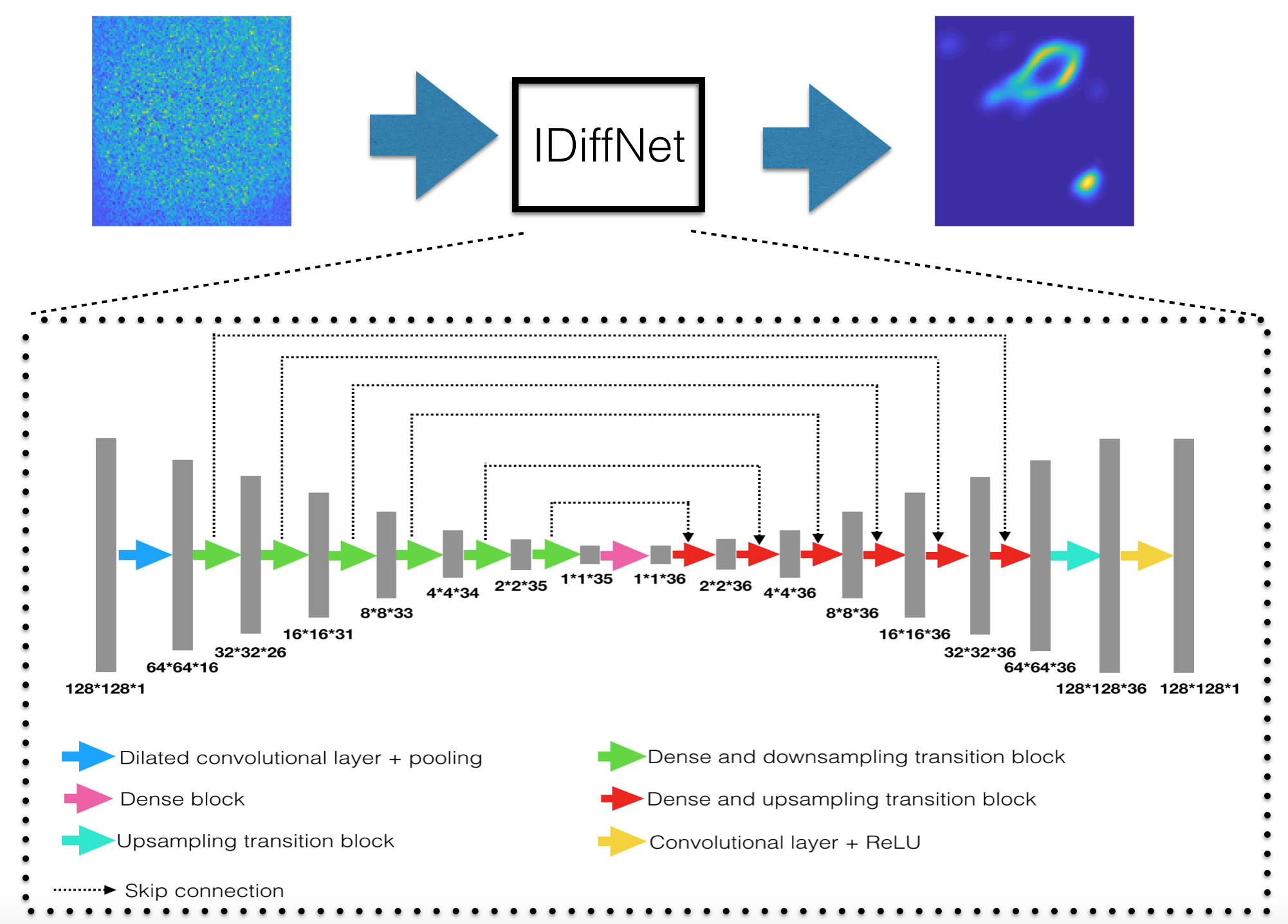}
\caption{IDiffNet, our densely connected neural network that images through diffuse media.}
\label{fig:dnn}
\end{figure*}

A diagram of IDiffNet is shown in Fig. \ref{fig:dnn}. The input to IDiffNet is the speckle pattern captured by the CMOS. It first passes through a dilated convolutional layer with filter size $5\times5$ and dilation rate 2 and is then successively decimated by 6 dense and downsampling transition blocks. After transmitting through another dense block, it successively passes through 6 dense and upsampling transition blocks and an additional upsampling transition layer. Finally, the signals pass through a standard convolutional layer with filter size $1\times1$ and the estimate of the object is produced. Due to the scattering by the glass diffusers, the intensity at one pixel of the image plane is influenced by several nearby pixels at the object plane. Therefore, we use dilated convolutions in all our dense blocks so as to increase the receptive field of the convolution filters. In addition, we also use skip connections to pass high frequency information learnt in the initial layers down the network towards the output reconstruction. Additional details about the architecture and training of IDiffNet are provided in the Supplement, Section 3.

\section{Results and network analysis}
Our experiment consists two phases: training and testing.  During the training process, we randomly choose image samples from a training database. The space bandwidth product of the original images are all $128\times128$ and we magnify each image by a factor of 4 before uploading to the SLM. The corresponding speckle patterns are captured by the CMOS. As mentioned in Section 2, we only crop the central $512\times 512$ square of the CMOS. We further downsample the captured speckle patterns by a factor of 4 and subtract from them a reference speckle pattern, which is obtained by uploading to the SLM a uniform image with all pixels equal to zero. The purpose of this subtraction operation is to eliminate the background noise on the CMOS and also to better extract differences between speckle patterns resulting from different objects. 

After the subtraction operation, we feed the resulting speckle patterns into IDiffNet for training. In this way, the input and output signal dimensions are both $128\times128$. We collected data from six separate experiment runs: each time we used training inputs from one of the three different databases: Faces-LFW \cite{huang2007labeled}, ImageNet \cite{russakovsky2015imagenet} or MNIST \cite{lecun2010mnist} and inserted one of the two glass diffusers that we have into the imaging system. Each of our training dataset consists of 10,000 object-speckle pattern pairs. These data were used to train six separate IDiffNets for evaluation. In the testing process,  we sample disjoint examples from the same database (Faces-LFW, ImageNet or MNIST) and other databases such as Characters, CIFAR \cite{krizhevsky2009learning} and Faces-ATT \cite{citeulike:2604432}. We upload these test examples to the SLM and capture their corresponding speckle patterns using the same glass diffuser as the training phase. We then input these speckle patterns to our trained IDiffNet and compare the output to the ground truth.

In training the IDiffNets, we use two different loss functions and compare their performances. The first loss function that we consider is the mean absolute error (MAE), is defined as:
\begin{equation}
\text{MAE}=\frac{1}{wh}\sum_{i=1}^{w}\sum_{j=1}^{h}\lvert Y(i,j)-G(i,j)\lvert
\end{equation}
Here, $w,h$ are the width and height of the output, $Y$ is the output of the last layer, and $G$ is the ground truth.

\begin{figure}[h!]
\centering\includegraphics[width=1\linewidth]{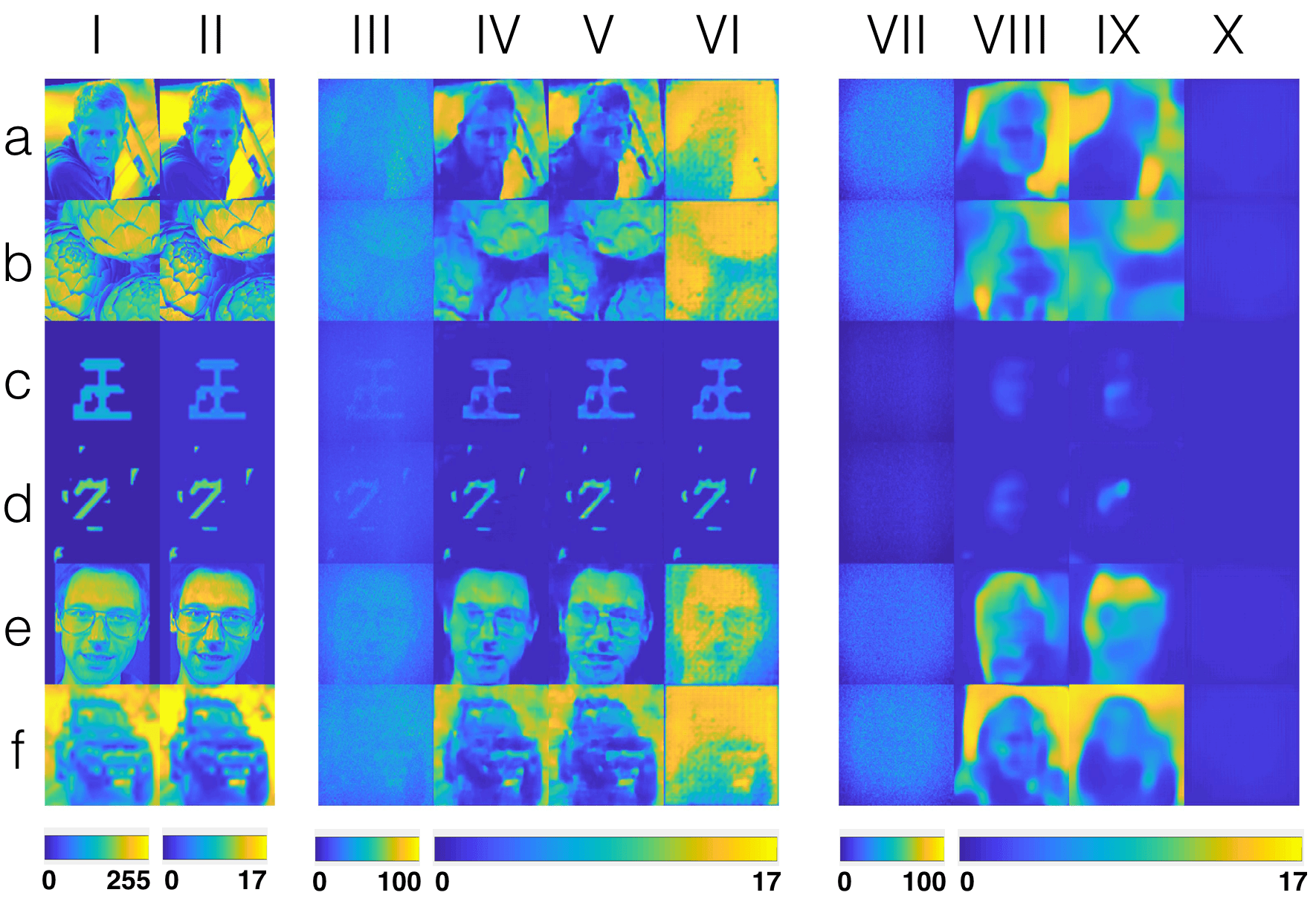}
\caption{Qualitative analysis of IDiffNet trained using MAE as the loss function. (i) Ground truth pixel value inputs to the SLM. (ii) Corresponding intensity images calibrated by SLM response curve. (iii) Raw intensity images captured by CMOS detector for 600-grit glass diffuser. (iv) IDiffNet reconstruction from raw images when trained using Faces-LFW dataset \protect{\cite{huang2007labeled}. (v) IDiffNet reconstruction when trained used ImageNet dataset \cite{russakovsky2015imagenet}}. (vi) IDiffNet reconstruction when trained used MNIST dataset \protect{\cite{lecun2010mnist}}. Columns (vii-x) follow the same sequence as (iii-vi) but in these sets the diffuser used is 220-grit.  Rows (a-f) correspond to the dataset from which the test image is drawn: (a) Faces-LFW, (b) ImageNet, (c) Characters, (d) MNIST, (e) Faces-ATT \protect{\cite{citeulike:2604432}}, (f) CIFAR \protect{\cite{krizhevsky2009learning}}, respectively. }
\label{fig:qual_MAE}
\end{figure}

\begin{figure}[h!]
\centering\includegraphics[width=1\linewidth]{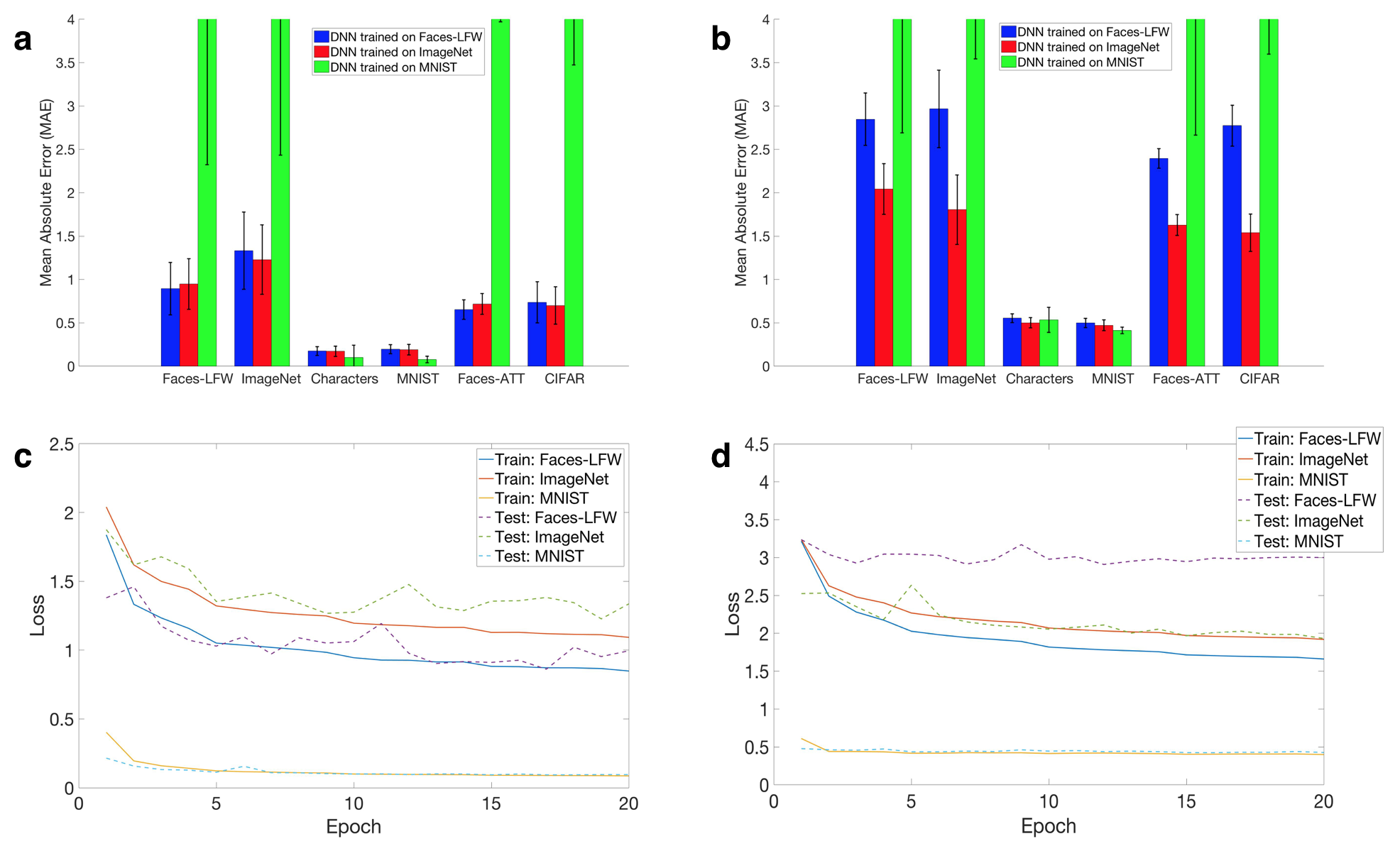}
\caption{Quantitative analysis of IDiffNet trained using MAE as the loss function. Test errors for IDiffNet trained on Faces-LFW (blue), ImageNet (red) and MNIST (green) on six datasets when the diffuser used is (a) 660-grit and (b) 220-grit. The training and testing error curves when the diffuser used is (c) 660-grit and (d) 220-grit. }
\label{fig:quan_MAE}
\end{figure}

The qualitative and quantitative reconstruction results when using MAE as the loss function are shown in Fig. \ref{fig:qual_MAE} and \ref{fig:quan_MAE}, respectively. From Fig. \ref{fig:qual_MAE}, we find that, generally speaking, IDiffNet's reconstruction performance for the 600-grit diffuser is better than that for the 220-grit diffuser. High quality reconstructions are achieved for the 600-grit diffuser when IDiffNets are trained on Faces-LFW (column iv) and ImageNet (column v). For the 220-grit diffuser, the best reconstruction is obtained when IDiffNet is trained on the ImageNet database (column ix). The recovered images are close to the low-pass filtered version of the original image, where we can visualize the general shape (salient features) but the high frequency features are missing. This result is expected since the scattering caused by the 220-grit diffuser is much stronger than that of the 600-grit diffuser, as we had already deduced from Fig. \ref{fig:psf}. As a result, we can still visualize some features of the object in the raw intensity image captured in the 600-grit diffuser case. By contrast,  what we capture in the 220-grit diffuser case looks indistinguishable from pure speckle, without any object details visible. This means we should expect it to be much more difficult for IDiffNet to do the inversion.

Noticeable from Fig. \ref{fig:qual_MAE} is that when IDiffNet is trained on MNIST for the 220-grit diffuser (column x), all the reconstructions seem to be uniform. This is due to the fact that the objects that this IDiffNet was trained on were sparse; and, hence, it also tends to make sparse estimates. Unfortunately, in this case the sparse local minima where IDiffNet is trapped are featureless. Tackling this problem motivated us to examine the Negative Pearson Correlation Coefficient (NPCC) as alternative loss function.

The NPCC is defined as \cite{neto2013image}:
\begin{equation}
\text{NPCC}=\frac{-1\times\sum_{i=1}^{w}\sum_{j=1}^{h}(Y(i,j)-\tilde{Y})(G(i,j)-\tilde{G})}{\sqrt{\sum_{i=1}^{w}\sum_{j=1}^{h}(Y(i,j)-\tilde{Y})^2}\sqrt{\sum_{i=1}^{w}\sum_{j=1}^{h}(G(i,j)-\tilde{G})^2}}
\end{equation}
Here, $\tilde{G}$ and $\tilde{Y}$ are the mean value of the ground truth $G$ and the IDiffNet output $Y$, respectively.

\begin{figure}[h!]
\centering\includegraphics[width=1\linewidth]{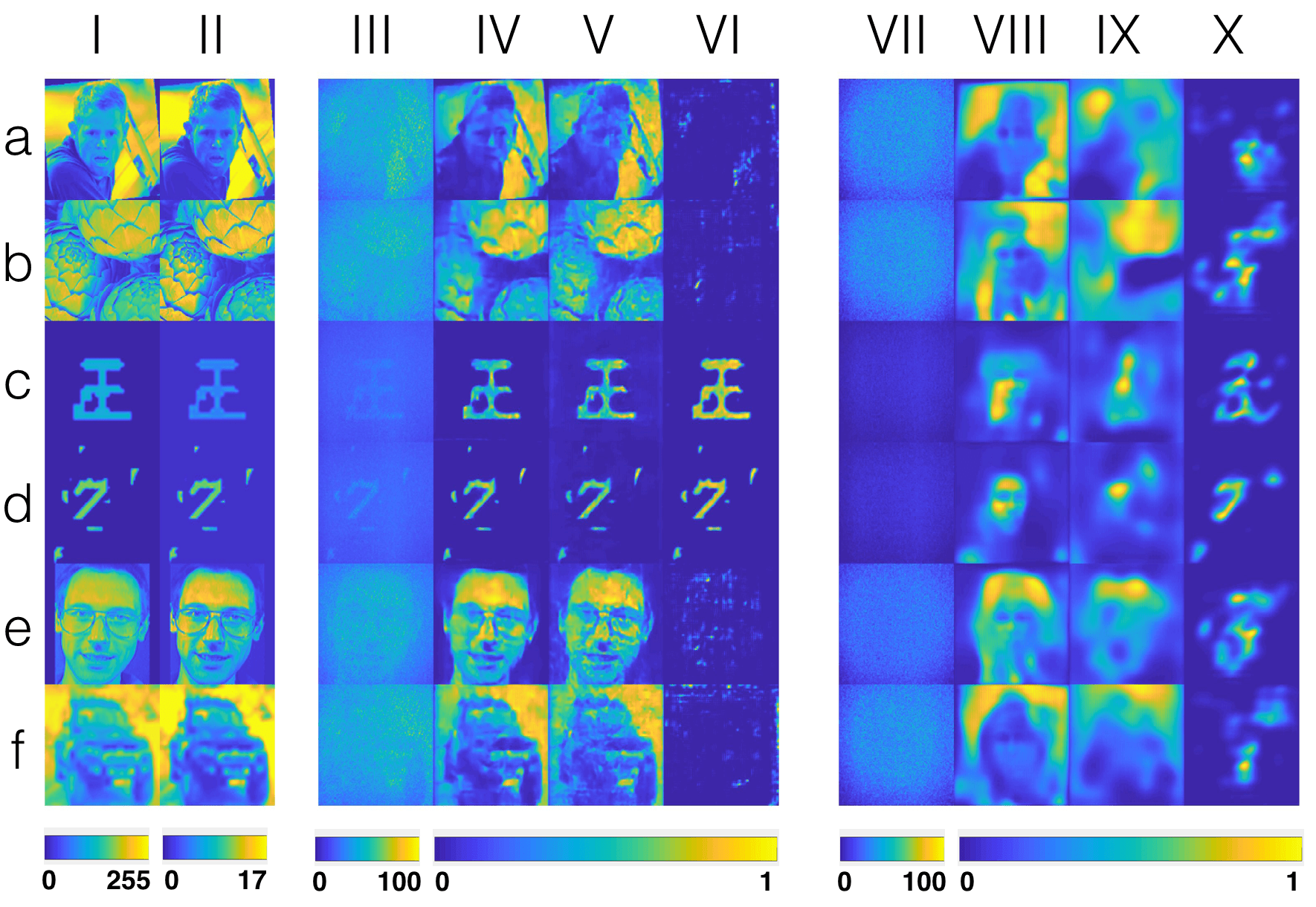}
\caption{Qualitative analysis of IDiffNets trained using NPCC as the loss function. (i) Ground truth pixel value inputs to the SLM. (ii) Corresponding intensity images calibrated by SLM response curve. (iii) Raw intensity images captured by CMOS detector for 600-grit glass diffuser. (iv) IDiffNet reconstruction from raw images when trained using Faces-LFW dataset \protect{\cite{huang2007labeled}}. (v) IDiffNet reconstruction when trained used ImageNet dataset \protect{\cite{russakovsky2015imagenet}}. (vi) IDiffNet reconstruction when trained used MNIST dataset \protect{\cite{lecun2010mnist}}. Columns (vii-x) follow the same sequence as (iii-vi) but in these sets the diffuser used is 220-grit.  Rows (a-f) correspond to the dataset from which the test image is drawn: (a) Faces-LFW, (b) ImageNet, (c) Characters, (d) MNIST, (e) Faces-ATT \protect{\cite{citeulike:2604432}}, (f) CIFAR \protect{\cite{krizhevsky2009learning}}, respectively.}
\label{fig:qual_PCC}
\end{figure}

\begin{figure}[h!]
\centering\includegraphics[width=1\linewidth]{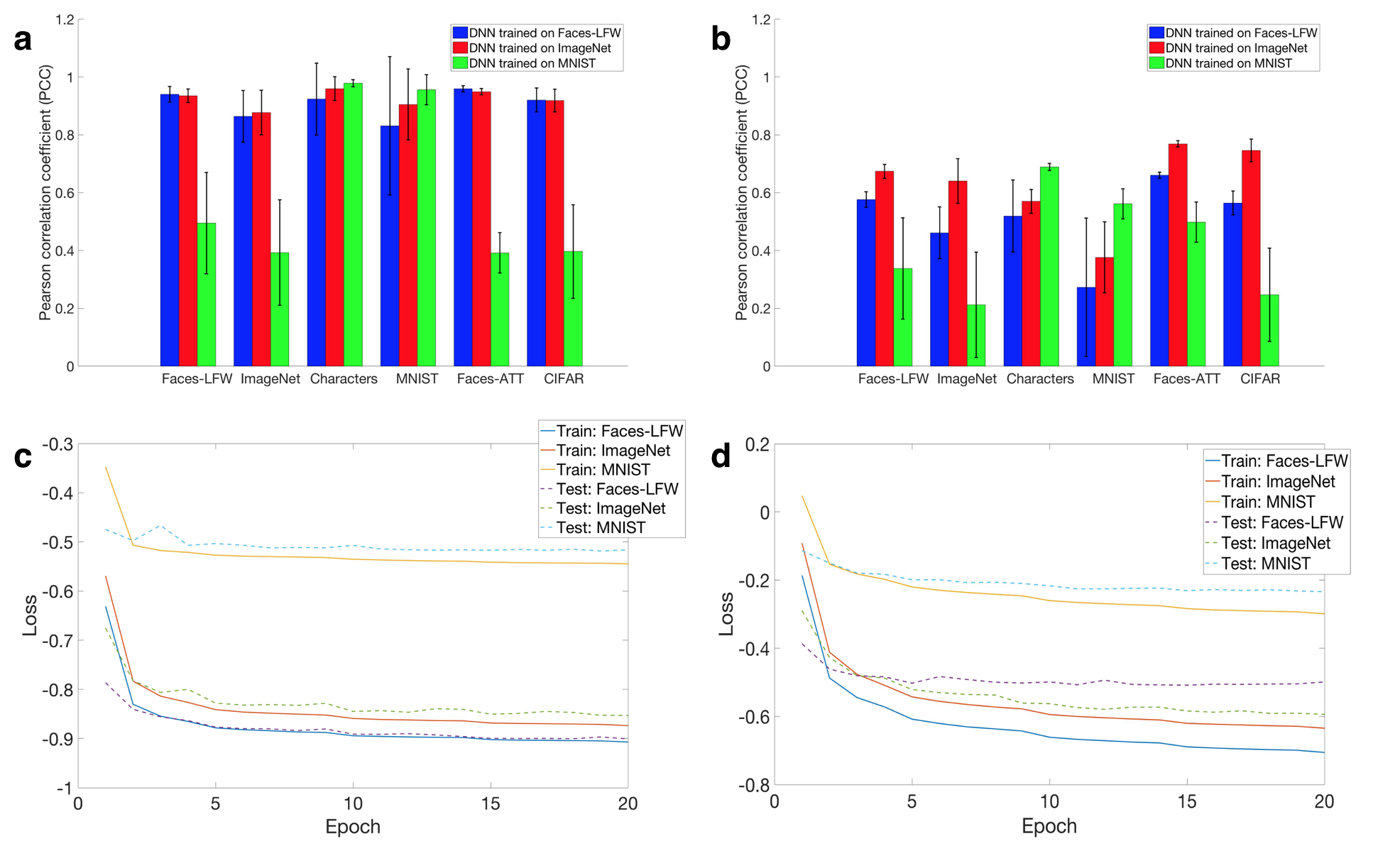}
\caption{Quantitative analysis of our trained deep neural networks for using NPCC as the loss function. Test errors for the IDiffNets trained on Faces-LFW (blue), ImageNet (red) and MNIST (green) on six datasets when the diffuser used is (a) 660-grit and (b) 220-grit. The training and testing error curves when the diffuser used is (c) 660-grit and (d) 220-grit.}
\label{fig:quan_PCC}
\end{figure}

The qualitative and quantitative reconstruction results using NPCC as the loss function are shown in Fig. \ref{fig:qual_PCC} and \ref{fig:quan_PCC}, respectively. The reconstructed images are normalized since the NPCC value will be the same if we multiply the reconstruction by any positive constants.  Similar to the case where MAE is used as the loss function, the reconstruction is better in the 600-grit diffuser case than the 220-grit diffuser case. However, when IDiffNet is trained on MNIST for the 220-grit diffuser (column x), high quality reconstruction is achieved for the test images comes from Characters and MNIST database (row c and d). This is in contrast to the MAE-trained case, thus indicating that NPCC is a more appropriate loss function to use in this case. It helps IDiffNet to learn the sparsity in the ground truth and in turn use the sparsity as a strong prior for estimating the inverse. In addition, when trained on ImageNet for the 220-grit diffuser (column ix), IDiffNet is still able to reconstruct the general shape (salient features) of the object. But the NPCC-trained reconstructions are visually slightly worse compared with the MAE-trained cases. 

In both MAE and NPCC training cases, IDiffNet performance also depends on the dataset that it is trained on. From Fig. \ref{fig:qual_MAE} and \ref{fig:qual_PCC}, we observe that IDiffNet generalizes best when being trained on ImageNet and has the most severe overfitting problem when being trained on MNIST. Specifically, when IDiffNet is trained on MNIST, even for the 600-grit diffuser (column vi), it works well if the test image comes from the same database or a database that shares the same sparse characteristics as  MNIST (e.g. characters). It gives much worse reconstruction when the test image comes from a much different database. When IDiffNet is trained on Faces-LFW, it generalizes well for the 600-grit diffuser, but for the 220-grit diffuser it exhibits overfitting: it tends to reconstruct a face at the central region, as Horisaki's case. When IDiffNet is trained on ImageNet, it generalizes well even for the 220-grit diffuser. As we can see in column ix, for all the test images, IDiffNet is able to at least reconstruct the general shapes (salient features) of the objects. This indicates that IDiffNet has learned at the very least a generalizable mapping of low-level textures between the captured speckle patterns and the input images. Similar observation may also be made from Fig. \ref{fig:quan_MAE} and \ref{fig:quan_PCC}. From subplots (a) and (b) in both figures, we notice that the IDiffNets trained on MNIST have much higher MAEs/lower PCCs when tested on other databases.  As shown in subplot (d), the IDiffNets trained on Faces-LFW have a large discrepancy between training and test error, while for IDiffNets trained on ImageNet, the training and testing curves converge to almost the same level. An explanation for this phenomenon is that all the images in MNIST or Faces-LFW databases share the same characteristics (eg. sparse, circular shape), imposing a strong prior on IDiffNet. On the other hand, the ImageNet database consists a mixture of generic images that not have too much in common. As a result, IDiffNet trained on ImageNet generalizes better. It is worth noting that overfitting in our case evidences itself as face-looking ``ghosts'' occurring when IDiffNet trained on Faces-LFW tries to reconstruct other kinds of images, for example (see Fig. \ref{fig:qual_PCC}, column viii). This is again similar to Horisaki's observations \cite{horisaki2016learning}.

From comparing the four possible combinations of weak {\it vs} strong scattering and constrained dataset (e.g. MNIST) {\it vs} generic dataset (e.g. ImageNet), we conclude the following: when scattering is weak, it is to our benefit to train the IDiffNets on a generic dataset because the resulting neural networks generalize better and can cope with the scattering also for general test images. On the other hand, when scattering is strong, it is beneficial to use a relatively constrained dataset with strong sparsity present in the typical objects: the resulting neural networks are then more prone to overfitting, but now this works to our benefit for overcoming strong scattering (at the cost, of course, of working only for test objects coming from the more restricted database.) The choice of optimization functional makes this tradeoff even starker: MAE apparently does not succeed in learning the strong sparsity even for MNIST datasets, whereas the NPCC does much better, even being capable of reconstructing test objects under the most severe scattering conditions (220-grit diffuser, column x in Fig. \ref{fig:qual_PCC}) as long as the objects are drawn from the sparse dataset. These observations are summarized in Table. \ref{tab-1}.

\begin{table*}[h!]
\centering
\caption{Summary of reconstruction results in different cases. [$\surd$: Visually recognizable; $\bullet$: Salient feature recognizable; $\times$: Visually unrecognizable.] }
\label{tab-1}
\begin{tabular}{|l|l|ll|l|l|}
\hline
                                                       &                  & \multicolumn{2}{c|}{600-grit}                & \multicolumn{2}{c|}{220-grit} \\ \hline
                                                       &  Training dataset               & \multicolumn{1}{l|}{Loss: MAE} & Loss: NPCC & Loss: MAE     & Loss: NPCC    \\ \hline
\multicolumn{1}{|c|}{\multirow{3}{*}{Test: Faces-LFW}} & Faces-LFW & \multicolumn{1}{l|}{$\surd$}     &      $\surd$      &   $\bullet$            &  $\times$             \\ \cline{2-6} 
\multicolumn{1}{|c|}{}                                 &  ImageNet  & \multicolumn{1}{l|}{$\surd$}          &  $\surd$          &  $\bullet$               & $\bullet$                \\ \cline{2-6} 
\multicolumn{1}{|c|}{}                                 &  MNIST     & \multicolumn{1}{l|}{ $\times$  }          &  $\times$             & $\times$                 &  $\times$                \\ \hline
\multirow{3}{*}{Test: ImageNet}                        &  Faces-LFW & \multicolumn{1}{l|}{$\surd$}          &  $\surd$          &  $\times$                &  $\times$                \\ \cline{2-6} 
                                                       &  ImageNet  & \multicolumn{1}{l|}{$\surd$}          &   $\surd$         &      $\bullet$           &  $\bullet$               \\ \cline{2-6} 
                                                       &  MNIST     & \multicolumn{1}{l|}{$\times$}          &  $\times$          &    $\times$           &   $\times$            \\ \hline
\multirow{3}{*}{Test: Characters}                      &  Faces-LFW & \multicolumn{1}{l|}{$\surd$}          &   $\surd$         &   $\times$            & $\times$              \\ \cline{2-6} 
                                                       &  ImageNet  & \multicolumn{1}{l|}{$\surd$}          &   $\surd$         &     $\bullet$            & $\bullet$                \\ \cline{2-6} 
                                                       &  MNIST     & \multicolumn{1}{l|}{$\surd$}          &   $\surd$         &  $\times$             & $\surd$              \\ \hline
\multirow{3}{*}{Test: MNIST}                           &  Faces-LFW & \multicolumn{1}{l|}{$\surd$}          &  $\surd$          &    $\times$           & $\times$             \\ \cline{2-6} 
                                                       &  ImageNet  & \multicolumn{1}{l|}{$\surd$}          &   $\surd$         &   $\bullet$              &   $\bullet$          \\ \cline{2-6} 
                                                       &  MNIST     & \multicolumn{1}{l|}{$\surd$}          &    $\surd$        &     $\times$           & $\surd$              \\ \hline
\multirow{3}{*}{Test: Faces-ATT}                       &  Faces-LFW & \multicolumn{1}{l|}{$\surd$}          &  $\surd$          &   $\times$             & $\times$                \\ \cline{2-6} 
                                                       &  ImageNet  & \multicolumn{1}{l|}{$\surd$}          &   $\surd$         &  $\bullet$               &  $\bullet$             \\ \cline{2-6} 
                                                       &  MNIST     & \multicolumn{1}{l|}{$\times$ }          & $\times$            &   $\times$             &   $\times$             \\ \hline
\multirow{3}{*}{Test: CIFAR}                           & Faces-LFW & \multicolumn{1}{l|}{$\surd$}          & $\surd$            &   $\times$             &    $\times$            \\ \cline{2-6} 
                                                       &  ImageNet  & \multicolumn{1}{l|}{$\surd$}          &   $\surd$         &   $\bullet$              & $\bullet$               \\ \cline{2-6} 
                                                       &  MNIST     & \multicolumn{1}{l|}{$\times$ }          &  $\times$           &  $\times$              &  $\times$              \\ \hline
\end{tabular}
\end{table*}

\section{Comparison with denoising neural networks}
To get a sense of what IDiffNets learn, we first compare their reconstruction results with that of a denoising neural network. Specifically, we use ImageNet as our training database. To each image in the training dataset, we simulate a noisy image using Poisson noise and make the peak signal-to-noise ratio (PSNR) of the resulting noisy image visually comparable to that of the corresponding speckle image captured using the 600-grit diffuser.  We use Poisson noise other than different kinds of noise such as Gaussian because Poisson noise is signal-dependent, similar to the diffuser case.  We then train a denoising neural network using those noisy images. For the denoising neural network, we implement the residual network architecture \cite{remez2017deep}. Finally, we feed the test speckle images captured using the 600-grit diffuser into this denoising neural network and compare the outputs with those reconstructed by IDiffNet (using MAE as the loss function).

\begin{figure}[h!]
\centering\includegraphics[width=1\linewidth]{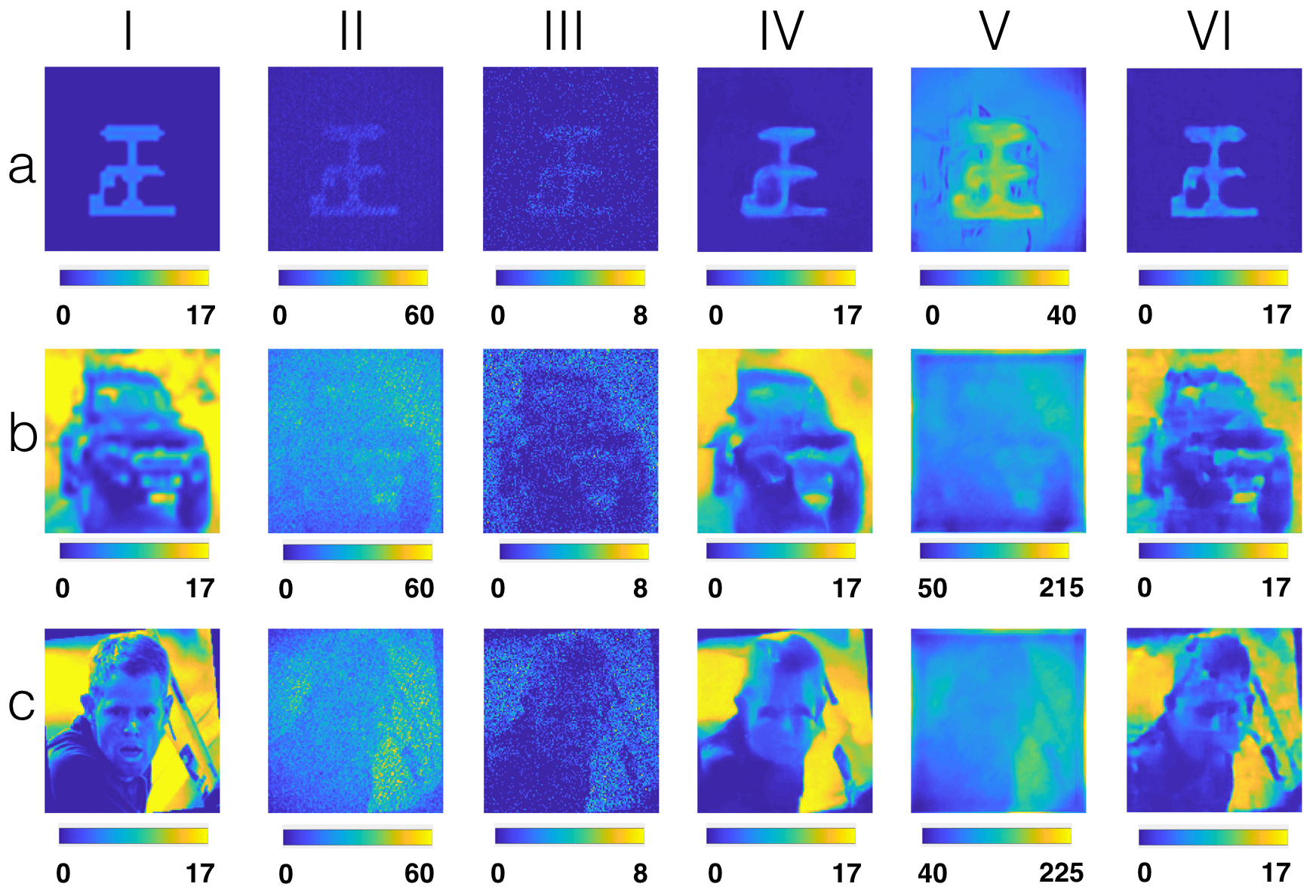}
\caption{Comparison between IDiffNets and a denoising neural network. (i) Ground truth intensity images calibrated by SLM response curve. (ii) Speckle images that we captured using the 600-grit diffuser (after subtracting the reference pattern). (iii) Noisy images generated by adding Poisson noise to the ground truth. (iv) Reconstructions of the denoising neural network when inputing the noisy image in (iii). (v) Reconstructions of the denoising neural network when inputing the speckle image in (ii). (vi) IDiffNet reconstructions when inputing the noisy image the speckle image in (ii). [The images shown in column vi are the same as those in the column v of Fig .\ref {fig:qual_MAE}, duplicated here for the readers' convenience]. Rows (a-c) correspond to the dataset from which the test image is drawn: (a) Characters, (b) CIFAR \cite{krizhevsky2009learning}, (c) Faces-LFW \cite{huang2007labeled}, respectively.}
\label{fig:compare}
\end{figure}

The comparison results are shown in Fig. \ref{fig:compare}. From column iv, we find that the denoising neural network works well when the test images are indeed noisy according to the Poisson model. However, as shown in column v, if we input the diffuse image into the denoising network, then it can only output a highly blurred image, much worse than IDiffNet given the same diffuse input, as can be seen in column vi. This result demonstrates that IDiffNet is not doing denoising, although the speckle image obtained using the 600-grit diffuser visually looks similar to a noisy image. We posit the reason for this as follows: Poisson noise operates pixel-wise. Consequently, denoising for Poisson noise is effectively another pixel-wise operation that does not depend much on spatial neighborhood, except to the degree that applying priors originating from the structure of the object helps to denoise severely affected signals. A denoising neural network, then, learns spatial structure only as a prior on the class of objects it is trained on. However, this is not the case when imaging through a diffuser: then every pixel value in the speckle image is influenced by a set of nearby pixels in the original image. This may also be seen from the PSF plots shown in Fig. \ref{fig:psf}. The denoising neural network fails because it has not learnt the spatial correlations between the nearby pixels and the correct kernel of our imaging system, as our IDiffNet has.

\begin{figure}[h!]
\centering\includegraphics[width=1\linewidth]{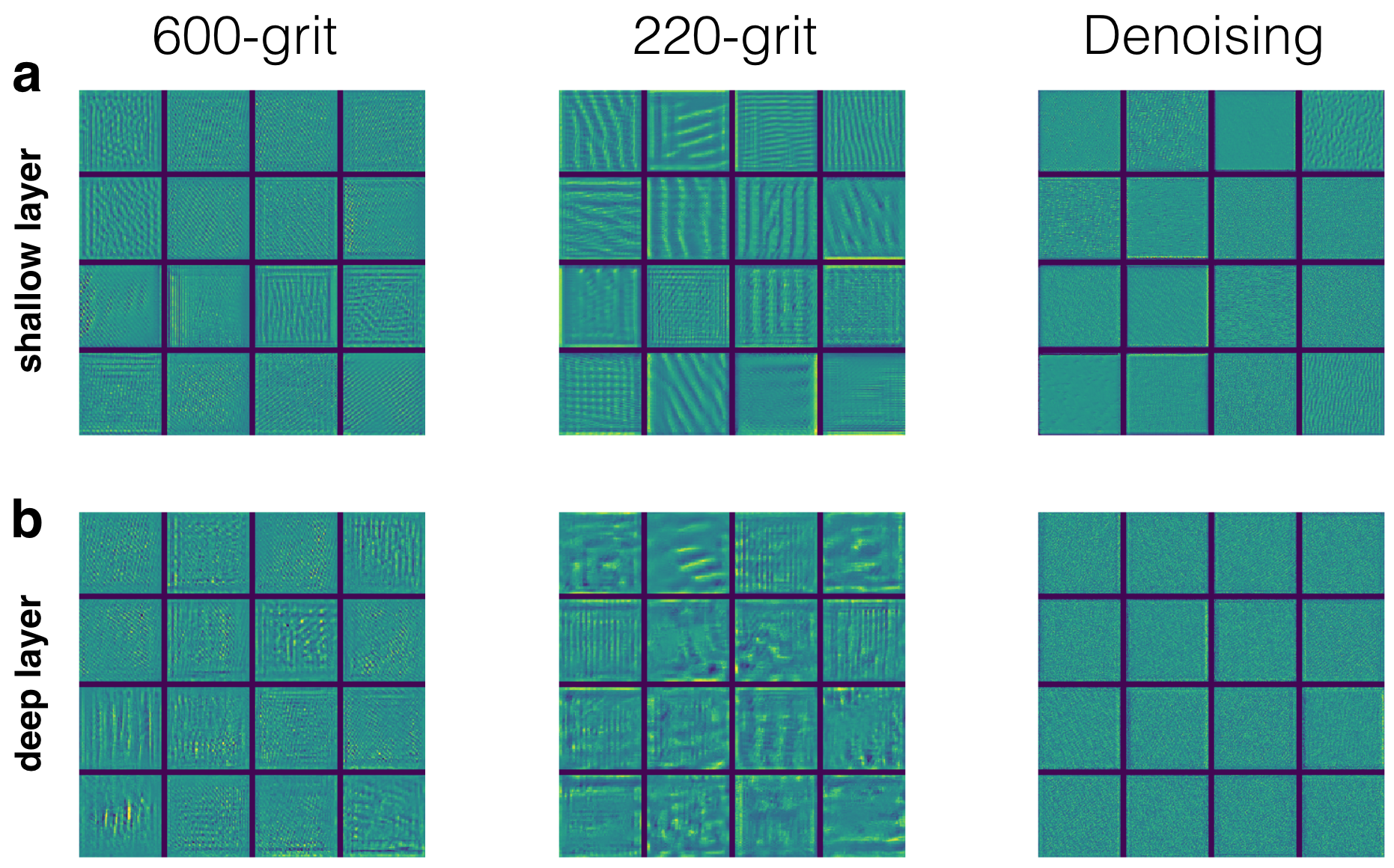}
\caption{Maximally-activated patterns (MAPs) for different DNNs. (a) $128\times 128$ inputs that maximally activate the filters in the convolutional layer at depth 5. (b) $128\times 128$ inputs that maximally activate the filters in the convolutional layer at depth 13. [There are actually more than 16 filters at each convolutional layer, but we only show the 16 filters have the highest activations here.]}
\label{fig:map}
\end{figure}

We also examined the maximally-activated patterns (MAPs) of the IDiffNets and the denoising neural network; {\it i.e.} what types of inputs would maximize network filter response (gradient descent on the input with average filter response as loss function) \cite{zeiler2014visualizing}.  Fig. \ref{fig:map} shows the MAP analysis of two convolutional layers at different depths for all the three neural networks. For both the shallow and deep layers, we find the MAPs of our IDiffNets are qualitatively different from those of the denoising network. This further corroborates that IDiffNet is not merely doing denoising. In addition, the MAPs of the 600-grit IDiffNet show finer textures compared with that of the 220-grit IDiffNet, indicating that the IDiffNet learns better in the 600-grit diffuser case.

\section{Conclusions}

We have demonstrated that IDiffNets, built according to the densely connected convolutional neural network architecture, can be used as an end to end approach for imaging through scattering media. The reconstruction performance depends on the scattering strength of the diffusers, the type of the training dataset (in particular, its sparsity), as well as the loss function used for optimization. The IDiffNets seem to learn automatically both the properties of the scattering media, as well as the priors restricting the objects where the network is supposed to perform well, depending on what the network was trained on. 

\section*{Funding Information}
 This research was funded by the Singapore National Research Foundation through the SMART program (Singapore-MIT Alliance for Research and Technology) and by the Intelligence Advanced Research Projects Agency (iARPA) through the RAVEN Program. Justin Lee acknowledges funding from the U.S. Department of Energy Computational Science Graduate Fellowship (CSGF) (DE-FG02-97ER25308).

\section*{}
\noindent
See Supplement for supporting content.

\bibliography{reference}

\begin{thebibliography}{10}
\newcommand{\enquote}[1]{``#1''}

\bibitem{tatarski2016wave}
V.~I. Tatarski, \emph{Wave propagation in a turbulent medium} (Courier Dover
  Publications, 2016).

\bibitem{ishimaru1978wave}
A.~Ishimaru, \emph{Wave propagation and scattering in random media}, vol.~2
  (Academic press New York, 1978).

\bibitem{denk1990two}
W.~Denk, J.~H. Strickler, W.~W. Webb \emph{et~al.}, \enquote{Two-photon laser
  scanning fluorescence microscopy,} Science \textbf{248}, 73--76 (1990).

\bibitem{moreaux2000membrane}
L.~Moreaux, O.~Sandre, and J.~Mertz, \enquote{Membrane imaging by
  second-harmonic generation microscopy,} JOSA B \textbf{17}, 1685--1694
  (2000).

\bibitem{hell1994breaking}
S.~W. Hell and J.~Wichmann, \enquote{Breaking the diffraction resolution limit
  by stimulated emission: stimulated-emission-depletion fluorescence
  microscopy,} Optics letters \textbf{19}, 780--782 (1994).

\bibitem{gustafsson2000surpassing}
M.~G. Gustafsson, \enquote{Surpassing the lateral resolution limit by a factor
  of two using structured illumination microscopy,} Journal of microscopy
  \textbf{198}, 82--87 (2000).

\bibitem{wilson2011optical}
T.~Wilson, \enquote{Optical sectioning in fluorescence microscopy,} Journal of
  microscopy \textbf{242}, 111--116 (2011).

\bibitem{lim2008wide}
D.~Lim, K.~K. Chu, and J.~Mertz, \enquote{Wide-field fluorescence sectioning
  with hybrid speckle and uniform-illumination microscopy,} Optics letters
  \textbf{33}, 1819--1821 (2008).

\bibitem{popoff2010measuring}
S.~Popoff, G.~Lerosey, R.~Carminati, M.~Fink, A.~Boccara, and S.~Gigan,
  \enquote{Measuring the transmission matrix in optics: an approach to the
  study and control of light propagation in disordered media,} Physical review
  letters \textbf{104}, 100601 (2010).

\bibitem{popoff2010image}
S.~Popoff, G.~Lerosey, M.~Fink, A.~Boccara, and S.~Gigan, \enquote{Image
  transmission through an opaque material,} Nature Communications \textbf{1},
  81 (2010).

\bibitem{dremeau2015reference}
A.~Dr{\'e}meau, A.~Liutkus, D.~Martina, O.~Katz, C.~Sch{\"u}lke, F.~Krzakala,
  S.~Gigan, and L.~Daudet, \enquote{Reference-less measurement of the
  transmission matrix of a highly scattering material using a dmd and phase
  retrieval techniques,} Optics express \textbf{23}, 11898--11911 (2015).

\bibitem{bertolotti2012non}
J.~Bertolotti, E.~G. van Putten, C.~Blum, A.~Lagendijk, W.~L. Vos, and A.~P.
  Mosk, \enquote{Non-invasive imaging through opaque scattering layers,} Nature
  \textbf{491}, 232--234 (2012).

\bibitem{katz2014non}
O.~Katz, P.~Heidmann, M.~Fink, and S.~Gigan, \enquote{Non-invasive single-shot
  imaging through scattering layers and around corners via speckle
  correlations,} Nature photonics \textbf{8}, 784--790 (2014).

\bibitem{stasio2016calibration}
N.~Stasio, C.~Moser, and D.~Psaltis, \enquote{Calibration-free imaging through
  a multicore fiber using speckle scanning microscopy,} Optics letters
  \textbf{41}, 3078--3081 (2016).

\bibitem{porat2016widefield}
A.~Porat, E.~R. Andresen, H.~Rigneault, D.~Oron, S.~Gigan, and O.~Katz,
  \enquote{Widefield lensless imaging through a fiber bundle via speckle
  correlations,} Optics express \textbf{24}, 16835--16855 (2016).

\bibitem{feng1988correlations}
S.~Feng, C.~Kane, P.~A. Lee, and A.~D. Stone, \enquote{Correlations and
  fluctuations of coherent wave transmission through disordered media,}
  Physical review letters \textbf{61}, 834 (1988).

\bibitem{freund1988memory}
I.~Freund, M.~Rosenbluh, and S.~Feng, \enquote{Memory effects in propagation of
  optical waves through disordered media,} Physical review letters \textbf{61},
  2328 (1988).

\bibitem{akkermans2007mesoscopic}
E.~Akkermans and G.~Montambaux, \emph{Mesoscopic physics of electrons and
  photons} (Cambridge university press, 2007).

\bibitem{gerchberg1972practical}
R.~W. Gerchberg, \enquote{A practical algorithm for the determination of the
  phase from image and diffraction plane pictures,} Optik \textbf{35}, 237--246
  (1972).

\bibitem{fienup1978reconstruction}
J.~R. Fienup, \enquote{Reconstruction of an object from the modulus of its
  fourier transform,} Optics letters \textbf{3}, 27--29 (1978).

\bibitem{grenander1993general}
U.~Grenander, \emph{General pattern theory-A mathematical study of regular
  structures} (Clarendon Press, 1993).

\bibitem{candes2006robust}
E.~J. Cand{\`e}s, J.~Romberg, and T.~Tao, \enquote{Robust uncertainty
  principles: Exact signal reconstruction from highly incomplete frequency
  information,} IEEE Transactions on information theory \textbf{52}, 489--509
  (2006).

\bibitem{brady2009compressive}
D.~J. Brady, K.~Choi, D.~L. Marks, R.~Horisaki, and S.~Lim,
  \enquote{Compressive holography,} Optics express \textbf{17}, 13040--13049
  (2009).

\bibitem{liu20153d}
H.-Y. Liu, E.~Jonas, L.~Tian, J.~Zhong, B.~Recht, and L.~Waller, \enquote{3d
  imaging in volumetric scattering media using phase-space measurements,}
  Optics express \textbf{23}, 14461--14471 (2015).

\bibitem{horisaki2016learning}
R.~Horisaki, R.~Takagi, and J.~Tanida, \enquote{Learning-based imaging through
  scattering media,} Optics express \textbf{24}, 13738--13743 (2016).

\bibitem{lyu2017exploit}
M.~Lyu, H.~Wang, G.~Li, and G.~Situ, \enquote{Exploit imaging through opaque
  wall via deep learning,} arXiv preprint arXiv:1708.07881  (2017).

\bibitem{Sinha:17}
A.~Sinha, J.~Lee, S.~Li, and G.~Barbastathis, \enquote{Lensless computational
  imaging through deep learning,} Optica \textbf{4}, 1117--1125 (2017).

\bibitem{rivenson2017phase}
Y.~Rivenson, Y.~Zhang, H.~Gunaydin, D.~Teng, and A.~Ozcan, \enquote{Phase
  recovery and holographic image reconstruction using deep learning in neural
  networks,} arXiv preprint arXiv:1705.04286  (2017).

\bibitem{rivenson2017deep}
Y.~Rivenson, Z.~Gorocs, H.~Gunaydin, Y.~Zhang, H.~Wang, and A.~Ozcan,
  \enquote{Deep learning microscopy,} arXiv preprint arXiv:1705.04709  (2017).

\bibitem{jin2017deep}
K.~H. Jin, M.~T. McCann, E.~Froustey, and M.~Unser, \enquote{Deep convolutional
  neural network for inverse problems in imaging,} IEEE Transactions on Image
  Processing \textbf{26}, 4509--4522 (2017).

\bibitem{satat2017object}
G.~Satat, M.~Tancik, O.~Gupta, B.~Heshmat, and R.~Raskar, \enquote{Object
  classification through scattering media with deep learning on time resolved
  measurement,} Optics Express \textbf{25}, 17466--17479 (2017).

\bibitem{horstmeyer2017convolutional}
R.~Horstmeyer, R.~Y. Chen, B.~Kappes, and B.~Judkewitz, \enquote{Convolutional
  neural networks that teach microscopes how to image,} arXiv preprint
  arXiv:1709.07223  (2017).

\bibitem{lecun1995convolutional}
Y.~LeCun, Y.~Bengio \emph{et~al.}, \enquote{Convolutional networks for images,
  speech, and time series,} The handbook of brain theory and neural networks
  \textbf{3361}, 1995 (1995).

\bibitem{zeiler2014visualizing}
M.~D. Zeiler and R.~Fergus, \enquote{Visualizing and understanding
  convolutional networks,} in \enquote{European conference on computer vision,}
   (Springer, 2014), pp. 818--833.

\bibitem{goodman2005introduction}
J.~W. Goodman, \emph{Introduction to Fourier optics} (Roberts and Company
  Publishers, 2005).

\bibitem{antipa2016single}
N.~Antipa, S.~Necula, R.~Ng, and L.~Waller, \enquote{Single-shot
  diffuser-encoded light field imaging,} in \enquote{Computational Photography
  (ICCP), 2016 IEEE International Conference on,}  (IEEE, 2016), pp. 1--11.

\bibitem{grit}
\url{https://www.unc.edu/~rowlett/units/scales/grit.html}.

\bibitem{huang2016densely}
G.~Huang, Z.~Liu, K.~Q. Weinberger, and L.~van~der Maaten, \enquote{Densely
  connected convolutional networks,} arXiv preprint arXiv:1608.06993  (2016).

\bibitem{huang2007labeled}
G.~B. Huang, M.~Ramesh, T.~Berg, and E.~Learned-Miller, \enquote{Labeled faces
  in the wild: A database for studying face recognition in unconstrained
  environments,} Tech. rep., Technical Report 07-49, University of
  Massachusetts, Amherst (2007).

\bibitem{russakovsky2015imagenet}
O.~Russakovsky, J.~Deng, H.~Su, J.~Krause, S.~Satheesh, S.~Ma, Z.~Huang,
  A.~Karpathy, A.~Khosla, M.~Bernstein \emph{et~al.}, \enquote{Imagenet large
  scale visual recognition challenge,} International Journal of Computer Vision
  \textbf{115}, 211--252 (2015).

\bibitem{lecun2010mnist}
Y.~LeCun, C.~Cortes, and C.~J. Burges, \enquote{Mnist handwritten digit
  database,} AT\&T Labs [Online]. Available: http://yann. lecun. com/exdb/mnist
  \textbf{2} (2010).

\bibitem{krizhevsky2009learning}
A.~Krizhevsky and G.~Hinton, \enquote{Learning multiple layers of features from
  tiny images,} Tech. rep., University of Toronto (2009).

\bibitem{citeulike:2604432}
\enquote{At\&t database of faces,} Tech. rep., AT\&T Laboratories Cambridge.

\bibitem{neto2013image}
A.~M. Neto, A.~C. Victorino, I.~Fantoni, D.~E. Zampieri, J.~V. Ferreira, and
  D.~A. Lima, \enquote{Image processing using pearson's correlation
  coefficient: Applications on autonomous robotics,} in \enquote{Autonomous
  Robot Systems (Robotica), 2013 13th International Conference on,}  (IEEE,
  2013), pp. 1--6.

\bibitem{remez2017deep}
T.~Remez, O.~Litany, R.~Giryes, and A.~M. Bronstein, \enquote{Deep
  convolutional denoising of low-light images,} arXiv preprint arXiv:1701.01687
   (2017).

\end{thebibliography}

\end{document}